\documentclass[10pt,twocolumn,english,prd,superscriptaddress,nofootinbib,preprintnumbers,showpacs,floatfix]{revtex4-2}

\usepackage[utf8]{inputenc}
\usepackage{amsmath}
\usepackage{amsfonts}
\usepackage{amssymb}
\usepackage{float}
\usepackage{graphicx} 
\usepackage[caption=false]{subfig}
\graphicspath{ {figures/} }
\usepackage[usenames,dvipsnames]{xcolor}
\usepackage{hyperref}   
\hypersetup{
    colorlinks=true, 
    citecolor=MidnightBlue,
    linkcolor=MidnightBlue,
    urlcolor=Cyan}
\usepackage{tikz}
\usepackage{subcaption}
\usepackage{orcidlink}
\usepackage{verbatim} 
\usepackage{soul} 
\definecolor{purple}{rgb}{1,0,1}
\definecolor{lime}{HTML}{A6CE39} 

\newcommand{\orcidicon}{%
	\begin{tikzpicture}
	\draw[lime, fill=lime] (0,0) 
		circle [radius=0.16] 
		node[white] {{\fontfamily{qag}\selectfont \tiny ID}};
	\draw[white, fill=white] (-0.0625,0.095) 
		circle [radius=0.007];
	\end{tikzpicture}	\hspace{-2mm}
}
\newcommand\orcidEdnaldo{{\href{https://orcid.org/0000-0002-9388-8373}{\orcidicon}}}
\newcommand\orcidFrancisco{{\href{https://orcid.org/0000-0002-9388-8373}{\orcidicon}}}
\newcommand\orcidManuel{{\href{https://orcid.org/0000-0001-8586-0285}{\orcidicon}}}
\newcommand\orcidTarciso{{\href{https://orcid.org/0009-0007-0450-2672}{\orcidicon}}}
\newcommand\orcidHenrique{{\href{https://orcid.org/0000-0001-7565-4277}{\orcidicon}}}
\newcommand\orcidLuis{{\href{https://orcid.org/0009-0009-4322-6484}{\orcidicon}}}
\newcommand\orcidDaniela{{\href{https://orcid.org/0009-0002-3463-0417}{\orcidicon}}}
\newcommand\orcidJorde{{\href{https://orcid.org/0009-0001-3344-2986}{\orcidicon}}}

\newcommand\orcidDiego{{\href{https://orcid.org/0000-0003-3984-9864}{\orcidicon}}}
\begin{document}

\title{Regular Bardeen black holes within non-minimal scalar-linear electrodynamic couplings}

\author{Daniela S. J. Cordeiro\orcidDaniela\!\!}         \email{fc52853@alunos.ciencias.ulisboa.pt}        
\affiliation{Instituto de Astrof\'{i}sica e Ci\^{e}ncias do Espa\c{c}o, Faculdade de Ci\^{e}ncias da Universidade de Lisboa, Edifício C8, Campo Grande, P-1749-016 Lisbon, Portugal}

	\author{Ednaldo L. B. Junior\orcidEdnaldo\!\!} \email{ednaldobarrosjr@gmail.com}
\affiliation{Faculdade de F\'{i}sica, Universidade Federal do Pará, Campus Universitário de Tucuruí, CEP: 68464-000, Tucuruí, Pará, Brazil}
\affiliation{Programa de P\'{o}s-Gradua\c{c}\~{a}o em F\'{i}sica, Universidade Federal do Sul e Sudeste do Par\'{a}, 68500-000, Marab\'{a}, Par\'{a}, Brazill}

	\author{Jos\'{e} Tarciso S. S. Junior\orcidTarciso\!\!}
 \email{tarcisojunior17@gmail.com}
\affiliation{Faculdade de F\'{\i}sica, Programa de P\'{o}s-Gradua\c{c}\~{a}o em 
F\'isica, Universidade Federal do 
 Par\'{a},  66075-110, Bel\'{e}m, Par\'{a}, Brazil}

	\author{Francisco S. N. Lobo\orcidFrancisco\!\!} \email{fslobo@ciencias.ulisboa.pt}
\affiliation{Instituto de Astrof\'{i}sica e Ci\^{e}ncias do Espa\c{c}o, Faculdade de Ci\^{e}ncias da Universidade de Lisboa, Edifício C8, Campo Grande, P-1749-016 Lisbon, Portugal}
\affiliation{Departamento de F\'{i}sica, Faculdade de Ci\^{e}ncias da Universidade de Lisboa, Edif\'{i}cio C8, Campo Grande, P-1749-016 Lisbon, Portugal}

	\author{Jorde A. A. Ramos\orcidJorde\!\!}
 \email{jordealves@ufpa.br}
\affiliation{Faculdade de F\'{\i}sica, Programa de P\'{o}s-Gradua\c{c}\~{a}o em 
F\'isica, Universidade Federal do 
 Par\'{a},  66075-110, Bel\'{e}m, Par\'{a}, Brazil}
 
	\author{Manuel E. Rodrigues\orcidManuel\!\!}
	\email{esialg@gmail.com}
	\affiliation{Faculdade de F\'{\i}sica, Programa de P\'{o}s-Gradua\c{c}\~{a}o em 
F\'isica, Universidade Federal do 
 Par\'{a},  66075-110, Bel\'{e}m, Par\'{a}, Brazil}
\affiliation{Faculdade de Ci\^{e}ncias Exatas e Tecnologia, 
Universidade Federal do Par\'{a}\\
Campus Universit\'{a}rio de Abaetetuba, 68440-000, Abaetetuba, Par\'{a}, 
Brazil}

\author{Diego Rubiera-Garcia\orcidDiego\!\!} \email{ drubiera@ucm.es}
\affiliation{Departamento de Física Téorica and IPARCOS, Universidad Complutense de Madrid, E-28040 Madrid, Spain}

\author{Luís F. Dias da Silva\orcidLuis\!\!} 
        \email{fc53497@alunos.fc.ul.pt}
\affiliation{Instituto de Astrof\'{i}sica e Ci\^{e}ncias do Espa\c{c}o, Faculdade de Ci\^{e}ncias da Universidade de Lisboa, Edifício C8, Campo Grande, P-1749-016 Lisbon, Portugal}


 \author{Henrique A. Vieira\orcidHenrique\!\!} \email{henriquefisica2017@gmail.com}
\affiliation{Faculdade de F\'{i}sica, Programa de P\'{o}s-Gradua\c{c}\~{a}o em F\'{i}sica, Universidade Federal do Par\'{a}, 66075-110, Bel\'{e}m, Par\'{a}, Brazil}


\begin{abstract}
	In this paper, we demonstrate that a linear electromagnetic matter source can be constructed for regular black hole solutions, illustrated by the case of the Bardeen black hole. Our method relies on coupling a scalar field and its potential to General Relativity (GR) alongside electrodynamics. Both fields are minimally coupled to GR but interact with each other. We validate this approach by interpreting the Bardeen solution either as  a purely magnetic or as a purely electric configuration for a spherically symmetric metric with an arbitrary electrodynamics Lagrangian $\mathcal{L}(F)$ and a coupling function $W(\varphi)$. By choosing an appropriate form of $W$, the electromagnetic Lagrangian can be rendered linear. We then analyze the behavior of the coupling function and the scalar field potential, emphasizing their relation to the magnetic or electric charge. Finally, we compare the advantages and disadvantages of this framework with the conventional formalism, in which regular charged black holes are supported solely by nonlinear electrodynamics. 
\end{abstract}

\date{\today}

\maketitle

\section{Introduction}

The existence of black holes is a central prediction not only of Einstein's General Relativity (GR) \cite{Einstein2,Einstein1905}, but also of many alternative theories of gravity formulated through geometrical constructions \cite{Capozziello:2011et}. Within GR, however, black holes inevitably exhibit a problematic feature that threatens the consistency of the theory itself: the presence of a space-time singularity at their cores. Singularities are fundamentally linked to the breakdown of geodesic completeness, whereby the trajectory of an observer (or a light ray) terminates after, or originates before, a finite interval of affine time \cite{Senovilla:2014gza}.  
In such geometrical theories of gravity, pathological behavior can likewise be traced back to the existence of curvature singularities, namely the divergence of certain curvature scalars as the affine time approaches a specific value \cite{Rubin,Inverno}. A solution can therefore be classified as \textit{regular} if it is free from singularities according to these criteria \cite{Bambi,Lan:2023cvz}.

Historically, the Bardeen solution \cite{Bardeen1} was the first to achieve this goal. It was originally introduced as a toy model, without an associated Lagrangian or matter source to support it. This posed the challenge of turning it into an exact solution of Einstein’s equations \cite{Borde:1996df,Cabo:1997rm,Mars:1996khm,Barrabes:1995nk}. This challenge was eventually resolved by Ayón-Beato and García \cite{Beato}, who interpreted Bardeen’s model as an exact, static, and spherically symmetric solution of Einstein’s equations sourced by a self-gravitating magnetic charge within the framework of nonlinear electrodynamics (NED).  
The NED framework was originally introduced by Born and Infeld in 1934 \cite{Born:1933pep,Born:1934gh} as a modification of Maxwell’s classical electrodynamics. Their theory aimed to remove singularities inherent to Maxwell’s framework—such as the divergent self-energy of point-like charges—by imposing a finite upper bound on the field strength. Following this seminal contribution, several important developments further advanced the field. Notable examples include the Euler–Heisenberg model, which incorporates effective quantum corrections in the presence of strong fields \cite{Heisenberg:1936nmg}, and Plebanski’s formulation of a broader mathematical structure for NED \cite{Plebanski}. These foundational efforts provided the basis for numerous modern developments in NED, which continue to deepen our understanding of its implications for both classical and quantum physics \cite{Kruglov:2014hpa,Kruglov:2014iwa,Kruglov:2014iqa,Bandos:2020jsw}.

Following the pioneering work of Ayón-Beato and García, it was soon recognized that NEDs could be systematically employed to construct a broad class of regular black hole solutions \cite{regular4}. This approach has led to a wide variety of configurations, including electrically charged black holes \cite{regular2,regular5} as well as rotating solutions \cite{regular6,regular7}, thereby significantly enriching the landscape of regular space-times.  
Another remarkable development is the class of solutions known as black bounces (BB), first proposed by Simpson and Visser \cite{Simpson:2018tsi}. These solutions, which smoothly interpolate between black holes and traversable wormholes, have since been extended into numerous generalizations \cite{Lobo:2020ffi,Mazza:2021rgq,Xu:2021lff,Yang:2022ryf,Bronnikov:2022bud,Canate:2022gpy,Rodrigues2023,Pereira:2023lck,Lima:2023arg,Alencar:2024yvh,Bronnikov:2023aya,Lima:2022pvc} and have been extensively explored in terms of their phenomenological consequences \cite{Nascimento:2020ime,Cheng:2021hoc,Guerrero:2021ues,Jafarzade:2021umv,Jafarzade:2020ova,Jafarzade:2020ilt,Yang:2021cvh,Bambhaniya:2021ugr,Guo:2021wid}. In many of these models, the matter content sourcing the geometry combines NEDs with scalar fields \cite{Bronnikov:2021uta}, providing a richer structure than purely electromagnetic scenarios.  
Moreover, the construction of regular black holes within the NED framework has been extended beyond GR to modified theories of gravity, such as $f(R)$ gravity \cite{regular1,regular3,regular9}, $f(G)$ gravity \cite{regular9,regular8}, and rainbow gravity \cite{regular10}, among others.

Despite the considerable interest in constructing regular solutions within the NED framework, several important drawbacks remain. First, many proposed solutions, including the Bardeen model, do not recover the Maxwell limit in the weak-field regime. This is in conflict with astrophysical observations, which consistently indicate linear electromagnetic behavior at large distances, while potential signatures of NED effects are restricted to the vicinity of compact objects \cite{Mignani:2016fwz,Ejlli:2020yhk}. Second, the formulation of a fullly consistent framework for the thermodynamic laws and some associated relations becomes troublesome \cite{Rasheed:1997ns,Breton:2004qa,Gulin:2017ycu,Gonzalez:2009nn,Diaz-Alonso:2012lkh}. Third, from the semiclassical or quantum perspective, a consistent NED theory must satisfy the fundamental requirements of unitarity and causality, which are generally intertwined in quantum processes. However, when imposed on NEDs, it turns out that every regular black hole solution violates them at the regular center \cite{Shabad:2011hf,Bronnikov:2022ofk}. Fourth, for regular solutions supported by NEDs, it is also known that photons do not follow null geodesics of the background metric. Instead, they propagate along null geodesics of an effective metric \cite{Novello:1999pg,Toshmatov:2021fgm}. This effective description introduces several pathologies, such as the emergence of new singularities, the fact that any photon reaching them experiences infinite blueshift, and the spherical (areal) radius behaves as in a wormhole geometry \cite{Breton:2002td}. This is rather an obstacle for some descriptions, such as the description of shadows \cite{KumarWalia:2024yxn}. Finally, for a black hole solution to be physically viable, it must be stable under small perturbations of the metric. As a consequence, any NED supporting regular black hole solutions must satisfy additional constraints on the Lagrangian and its derivatives, as described in \cite{Bronnikov:2022ofk,Moreno:2002gg}. In summary, classical Maxwell theory remains the best established and most widely accepted electrodynamical framework, supported by more than a century of experimental confirmation. Its principal advantage over NEDs lies in its mathematical simplicity and phenomenological consistency across virtually all observational regimes.

The main aim of the present work is to demonstrate that it is possible to construct matter sources supporting regular black hole solutions that are described by linear Maxwell electrodynamics. The essential ingredient is the introduction of a non-minimal coupling between the electromagnetic field and a scalar field, while ensuring that both fields remain minimally coupled to GR. This approach provides a framework in which the electromagnetic sector preserves its linear structure, thus avoiding many of the difficulties typically associated with nonlinear electrodynamics. We illustrate the method by reinterpreting Bardeen’s solution within this framework, showing that it can be consistently described both as a purely magnetic and as a purely electric configuration. For completeness, we also provide a shorter reinterpretation of the Ayón-Beato and García solution along the same lines. The results presented here contribute to the broader discussion on how engineered solutions should be interpreted in terms of different combinations of gravitational and matter sources, as well as their potential theoretical and observational consequences , see e.g. \cite{Bronnikov:2021uta,Rodrigues:2023vtm,Bolokhov:2024sdy,Bronnikov:2024izh}.

The paper is organized as follows: In Sec.~\ref{sec:um}, we provide a brief overview of Bardeen’s solution and the standard interpretation in which NEDs are taken as the sole matter source. In Sec.~\ref{sec:dois}, we develop our formalism, first addressing the magnetic case in Sec.~\ref{sec:doisa} and then the electric case in Sec.~\ref{sec:doisb}, obtaining in both cases the Bardeen solutions as supported by a linear electromagnetic Lagrangian with distinct coupling functions. Section~\ref{sec:conclu} contains our discussion and concluding remarks. In Appendix~\ref{sec:ap1}, we consider the Ayón-Beato and García solution as an additional example of the formalism. Throughout the paper we adopt geometrized units ($G=1$, $c=1$) and the metric signature $(+,-,-,-)$.

\section{The standard Bardeen solution from a NED source \label{sec:um}}

Einstein’s equations lie at the core of GR, describing the interplay between the geometry of space-time and the distribution of matter and energy in the universe. They can be derived through the variational principle, starting from the action
\begin{equation}
    S = \int d^4x\sqrt{-g} \left[ R + 2 \kappa^2 \mathcal{L}(F)  \right].
    \label{açãoNED}
\end{equation}
This action corresponds to the standard Einstein–Hilbert action coupled to nonlinear electrodynamics, with the following definitions: $R$ denotes the scalar curvature, $g$ is the determinant of the metric $g_{\mu\nu}$, and $\kappa^2 = 8 \pi$ is Newton’s constant in geometrized units. The function $\mathcal{L}(F)$ represents the electrodynamics matter Lagrangian, which, in general, may be any function of the Maxwell invariant $F$, given by
\begin{equation} \label{eq:Fmunu0}
    F  = \frac{1}{4} F^{\mu \nu}F_{\mu \nu},
\end{equation}
where $F_{\mu \nu}$ is the field strength tensor tensor, built out of the  quadri-potential  $A_{\mu}$
\begin{equation} \label{eq:Fmunu}
    F_{\mu \nu} = \partial_{\mu} A_{\nu} - \partial_{\nu}A_{\mu}.
\end{equation}

Varying the action \eqref{açãoNED} with respect to the metric $g_{\mu \nu},$ we obtain Einstein's equations
\begin{equation}
    R_{\mu \nu} - \frac{1}{2} g_{\mu \nu}R = \kappa^2 T_{\mu \nu},
    \label{eqeinstein}
\end{equation}
where $R_{\mu \nu}$ is the Ricci tensor and $T_{\mu \nu}$ is the energy-momentum tensor, which for NEDs reads as
\begin{equation}
    T_{\mu \nu} = g_{\mu \nu}\mathcal{L}(F)- \mathcal{L}_FF_{\mu}^{\ \alpha}F_{\alpha \nu},
    \label{7}
\end{equation}
with $\mathcal{L}_F \equiv d\mathcal{L}/dF$.

Within this framework, Ayón-Beato and García \cite{Beato} demonstrated that the regular black hole solution originally proposed by Bardeen \cite{Bardeen1} can be interpreted as being sourced by a magnetically charged field described by a NED Lagrangian density \cite{Rodrigues3}
\begin{equation}
    \mathcal{L}(F) = \frac{3}{8 \pi sq_m^2 } \left(   \frac{\sqrt{2 q_m^2 F}}{2+\sqrt{2 q_m^2 F}} \right)^{5/2},
    \label{lbardeen}
\end{equation}
where $s = |q_m|/(2 m)$, and $q_m$ is the magnetic charge. For the sake of this work lets us briefly review the basis behind this derivation.

Let us consider a static and spherically symmetric space-time given by the line element
\begin{equation}
    ds^2 = A(r) dt^2 - \frac{1}{A(r)} dr^2 -r^2( d \theta^2 + \sin^2{\theta} d \phi^2),
\end{equation}
with
\begin{equation}
    A(r) = 1 - \frac{2 M(r)}{r},
    \label{frbardeen}
\end{equation}
where $M(r)$ is a arbitrary function of the radial coordinate. Now, varying the action \eqref{açãoNED} with respect to $A_{\mu}$, we can obtain the NED field equations, which read as
\begin{equation}
    \nabla_{\mu} \left( F^{\mu \nu} \mathcal{L}_F  \right)  \equiv \partial_{\mu} \left(  \sqrt{-g} F^{\mu \nu} \mathcal{L}_F       \right)=0 .
    \label{maxwell}
\end{equation}
Considering that the solution is purely magnetic, we have from Eq. \eqref{eq:Fmunu} that there is a single non-vanishing component of the field strength tensor which takes the form
\begin{equation}
    F_{23}(\theta) = q_m \sin{\theta},
    \label{F23bardeen}
\end{equation}
where $q_m$ emerges as an integration constant. We can now easily compute the field scalar in Eq. (\ref{eq:Fmunu0}) as $F(r) = q_m^2/ 2 r^4$ and then the Lagrangian density \eqref{lbardeen} becomes
\begin{equation}
    \mathcal{L}(r) = \frac{3}{8 \pi sq_m^2 } \left(   \frac{q_m^2}{r^2+q_m^2} \right)^{5/2}.
    \label{eq:LrBardeen}
\end{equation}
Then, Einstein's equations \eqref{eqeinstein} lead us to
\begin{equation}
    \frac{d M(r)}{dr} = \frac{3 m q_m^2 r^2}{\left(q_m^2+r^2\right)^{5/2}},
\end{equation}
where $m$ is an integration constant. Integrating now with respect to $r$, we find Bardeen's mass function
\begin{equation}
    M(r) = \frac{m r^3}{\left(q_m^2+r^2\right)^{3/2}}.
    \label{massabardeen}
\end{equation}
Next, using Eq. \eqref{massabardeen}, the function $A(r)$ becomes
\begin{equation} 
    A(r) = 1-\frac{2 m r^2}{\left(q_m^2+r^2\right)^{3/2}}.
    \label{eq:A(r)}
\end{equation}
Given the fact that in the asymptotic limit one has $A(r \to \infty) \to 1-2m/r + \mathcal{O}(1/r^2)$, then $m$ can be intepreted as the asymptotic (ADM) mass of the system. 

Bardeen's model merits in removing the presence of any curvature singularity, inherent to the canonical GR black holes (Schwarzschild, Reissner-Nordstr\"om and Kerr-Newman solutions). To check this, we analyze the behaviour of the Kretschmann scalar, defined as 
\begin{equation}
    K = R^{\mu \nu \alpha \beta}R_{\mu \nu \alpha \beta},
\end{equation}
where $R^{\mu \nu \alpha \beta}$ is the Riemann tensor, which for Bardeen's metric (\ref{eq:A(r)}) yields the result
\begin{equation}
    K = \frac{12 m^2 \left(-4 q_m^6 r^2+47 q_m^4 r^4-12 q_m^2 r^6+8 q_m^8+4
   r^8\right)}{\left(q_m^2+r^2\right)^7}.
   \label{kresbardeen}
\end{equation}
This scalar has the following asymptotic limits 
\begin{equation}
    \lim_{r \to 0} K = \frac{96 m^2}{ q_m^6}, \ \ \ \ \lim_{r \to \infty} K = \frac{48m^2}{r^6} \to 0.
\end{equation}
Since $K$ is finite for any value of $r$ (and the same happens for every other scalar made up of similar contractions of the Riemann tensor), this corresponds to an everywhere regular space-time.

There is another side to Bardeen's model worth mentioning, namely, the fact that it can also be obtained by considering a purely electrical configuration \cite{Rodrigues3}. To see this, we first note that in this case the single non-vanishing component of the field strength tensor has the following form
\begin{equation} \label{eq:Felec}
    F^{10} (r) =  \frac{q_e}{r^2 \mathcal{L}_F(F)},
\end{equation}
and the non-zero components of Einstein's equations in this case are
\begin{equation}
    \begin{aligned}
        & \frac{2 M^{\prime}(r)}{r^2} = \kappa^2 \left(\mathcal{L} + \frac{q_e^2}{r^4 \mathcal{L}_F} \right), \\
        & \frac{ M^{\prime \prime}(r)}{r} = \kappa^2 \mathcal{L}.
    \end{aligned}
\end{equation}

Considering Eq. \eqref{massabardeen}, we can solve the above equations to find the Lagrangian density and its derivative as
\begin{equation}
    \mathcal{L} (r)  = \frac{q_e^2m \left(  6q_e^2 - 9r^2\right)}{\kappa^2 \left(q_e^2 +r^2\right)^{7/2}},
\end{equation}
and
\begin{equation}
    \mathcal{L}_F(r) = \frac{\kappa^2 \left(q_e^2+r^2\right)^{7/2}}{15mr^6},
\end{equation}
respectively. Therefore, one can use the above equations combined with (\ref{eq:Felec}) in order to find expressions for $F(r)$ and of $\mathcal{L}(F)$. However, there is a drawback of this procedure in that, and unlike the magnetic case, the electric description of Bardeen’s solution may lead to a multivalued structure for the Lagrangian density as a function of the field strength invariant. This feature is a well-known feature of electrically charged NED models aimed at singularity removal \cite{Bronnikov:2000yz} and will be addressed in the following section.

It is important to note that, although in the summary presented in this section the magnetic and electric cases have been treated in a seemingly ``reverse'' order, the procedure used to obtain these solutions in the original works is essentially the same: one writes Einstein’s equations minimally coupled to a chosen NED, selects a type of source (purely electric or magnetic), and then substitutes Bardeen’s metric to determine the corresponding Lagrangian density. In the following section, we will adopt the same steps, but with a non-minimal coupling between electrodynamics and a scalar field, allowing us to obtain a linear electromagnetic Lagrangian density.

\section{The Bardeen solution with a linear electromagnetic source}\label{sec:dois}

Our starting point is to consider a modification to the action \eqref{açãoNED}, adding a scalar field to the matter term and coupling it to the electrodynamic part. Thus, the action will be
\begin{eqnarray}
 S &=& \int d^4x \, \sqrt{-g} \bigl[  R + 2 \kappa^2 \bigl(V(\varphi)-  \varepsilon(\varphi) \nabla^a \varphi \nabla_a \varphi 
 	\nonumber \\  
 	&& + W(\varphi) \mathcal{L}(F) \bigr)\bigr],
\label{eq:acao}
\end{eqnarray}
where  $\varphi$ represents the scalar field and $V(\varphi)$ its associated potential, $\varepsilon (\varphi)$ is a quantity that can be a function of the scalar field and its value defines the contribution of the latter as canonical or phantom,  and $W(\varphi)$ is a function of the scalar field that performs the non-minimal coupling between the scalar and electromagnetic fields, while both of them are minimally coupled to GR. Varying the above action with respect to the metric, we obtain Einstein's equations in this case as
\begin{equation}
G_{\mu \nu} = \kappa^2 \left( T^{(\varphi)}_{\mu \nu} + W(\varphi) T^{(\text{NED})}_{\mu \nu} \right),
\label{eq:Einstein}
\end{equation}
where $T^{(\text{NED})}_{\mu \nu}$ is defined by Eq.\;\eqref{7} and the energy-momentum tensor associated with the scalar field is given by
\begin{eqnarray}
\label{eq:tmunuphi}
T^{(\varphi)}_{\mu \nu} &=& 2 \varepsilon(\varphi) \nabla_{\mu} \varphi \nabla_{\nu} \varphi -  \varepsilon(\varphi) \nabla^{\alpha} \varphi \nabla_{\alpha} \varphi \, g_{\mu \nu} 
	\nonumber \\ 
&&+ V(\varphi) g_{\mu \nu}. 
\end{eqnarray}
The key ingredient of this framework lies precisely in the presence of the function $W(\varphi)$, which in general must be different from $W(\varphi) =1$ (the usual minimally coupled case between matter fields) to obtain the desidered result.

Now varying the action with respect to the scalar field \( \varphi \), we obtain
\begin{equation}
2 \varepsilon(\varphi) \Box \varphi +  \varepsilon'(\varphi) \nabla^a \varphi \nabla_a \varphi = - V'(\varphi) + W'(\varphi) \mathcal{L}(F),
\label{eq:campoescalar}
\end{equation}
 with $\Box \equiv \nabla^\mu \nabla_\mu$. Furthermore, varying the action with respect to the vector potential $A_{\mu}$, we obtain the electromagnetic equations in this case as 
\begin{equation}
\nabla_\mu \left( W(\varphi) \, \mathcal{L}_F \, F^{\mu\nu} \right) = 0.
\label{eq:maxwell}
\end{equation}

It is important to realize that, in this framework, the scalar field equations \eqref{eq:campoescalar} and  the electromagnetic ones  \eqref{eq:maxwell} are coupled. This fact will require new strategies to find the expressions of the Lagrangian density supporting the Bardeen solution, which we analyze next, splitting our discussion into the magnetic and electric charges.

\subsection{Magnetic Bardeen solution}\label{sec:doisa}

To find a source of matter described by linear electrodynamics for the magnetically charged Bardeen solution, we will consider as ansatz the static and spherically symmetric configuration given by 
\begin{equation}
ds^2=A(r)dt^2-B(r)dr^2- \Sigma^2(r)\, d\Omega^2,\label{eq:ds}
\end{equation}
and a purely magnetic field given by Eq. \eqref{F23bardeen}. We point out that, due to spherical symmetry, this line element could be cast via two independent functions only, via a suitable change of coordinates, but we shall keep the above form out for convenience of our analysis. Note also that, in principle, the functions $A(r), \ B(r)$ and $\Sigma(r)$ are arbitrary functions of the radial coordinate $r$.
With this information, we find from Eq. \eqref{eq:Einstein} the following equations
\begin{widetext}
\begin{equation}
    \frac{\Sigma (r) B'(r) \Sigma '(r)-B(r) \left(2 \Sigma (r) \Sigma ''(r)+\Sigma '(r)^2\right)+B(r)^2}{B(r)^2
   \Sigma (r)^2}-\kappa ^2 \left(\frac{ \varepsilon (r) \varphi '(r)^2}{B(r)}+\mathcal{L}(r) W(r)+ V(r)\right) = 0,
   \label{eq:00}
\end{equation}
\begin{equation}
\frac{\Sigma (r) A'(r) \Sigma '(r)+A(r) \left(\Sigma '(r)^2-B(r)\right)}{A(r) B(r) \Sigma
   (r)^2}+\kappa ^2 \left(V(r)-\frac{ \varepsilon (r) \varphi '(r)^2}{B(r)}+\mathcal{L}(r) W(r)\right)    =0,
   \label{eq:11}
\end{equation}
\begin{equation}
\begin{aligned}
\frac{A(r) B'(r) \left(\Sigma (r) A'(r)+2 A(r) \Sigma '(r)\right)+B(r) \left(\Sigma (r) \left(A'(r)^2-2
   A(r) A''(r)\right)-2 A(r) \left(A'(r) \Sigma '(r)+2 A(r) \Sigma ''(r)\right)\right)}{4 A(r)^2 B(r)^2
   \Sigma (r)} 
   		\\
   	-\kappa ^2 \left(\frac{ \varepsilon (r) \varphi '(r)^2}{B(r)}+W(r) \left(L(r)-\frac{q_m^2
   \mathcal{L}_F(r)}{\Sigma (r)^4}\right)+ V(r)\right)    =0.
   \end{aligned}
   \label{eq:22}
\end{equation}
Now, we solve Eqs. \eqref{eq:00} and \eqref{eq:22} to find $\mathcal{L}$ and $\mathcal{L}_F$ as
\begin{equation}
    \mathcal{L} = \frac{\Sigma (r) B'(r) \Sigma '(r)+B(r)^2 \left(1- \kappa ^2 \Sigma (r)^2 V(r)\right)-B(r) \left(\Sigma
   '(r)^2+ \Sigma (r) \left(2\Sigma ''(r)+\kappa ^2 \Sigma (r) \varepsilon (r) \varphi
   '(r)^2\right)\right)}{\kappa ^2 B(r)^2 \Sigma (r)^2 W(r)},
   \label{eq:L}
\end{equation}
\begin{equation}
\begin{aligned}
   \mathcal{L}_F = \frac{1}{4 \kappa ^2 q_m^2 A(r)^2
   B(r)^2 W(r)} \Biggl\{ \Sigma (r)^2 \Big[-B(r) \Sigma (r)^2 A'(r)^2 +A(r) \Sigma (r) \bigl(2 B(r) \Sigma (r) A''(r)+A'(r)
   \bigl(2 B(r) \Sigma '(r) 
   	\\ 
   	-\Sigma (r) B'(r)\bigr)\bigr)  +2 A(r)^2 \left(\Sigma (r) B'(r) \Sigma '(r)-2
   B(r) \left(\Sigma (r) \Sigma ''(r)+\Sigma '(r)^2\right)+2 B(r)^2\right)\Bigr] \Biggr\}.
      \end{aligned}
      \label{eq:LF}
\end{equation}
\end{widetext}
respectively. 

These functions, derived from Einstein’s equations, are independent of one another. Therefore, to select only solutions that are consistent with electrodynamics, it is necessary to verify that the self-consistency equation
\begin{equation}
    \mathcal{L}_F - \frac{\partial \mathcal{L}}{\partial r} \left( \frac{\partial F}{\partial r} \right)^{-1} = 0,
\end{equation}
is satisfied for any ansatz made upon the above equations. Using these expressions in \eqref{eq:11}, we find
\begin{equation}
    \varepsilon  \varphi '^2 = \frac{\Sigma '(r) \left(B(r) A'(r)+A(r) B'(r)\right)-2 A(r) B(r) \Sigma ''(r)}{2 \kappa ^2 A(r) B(r) \Sigma
   (r)}.
   \label{eq:epsilon}
\end{equation}

Up to this point, $A$, $B$, and $\Sigma$ remain arbitrary functions of the radial coordinate $r$. However, it is worth noting that the numerator on the right-hand side of Eq.~\eqref{eq:epsilon} vanishes if we impose the conditions $B = 1/A$ and $\Sigma'' = 0$. This corresponds to the case of interest in the present work. Under these conditions, there are two possibilities: either $\varepsilon = 0$ or $\varphi(r) = \text{const}$. The second option is not physically meaningful, as it would render the scalar field constant and, consequently, the coupling function $W(\varphi)$ trivial. Therefore, we set $\varepsilon = 0$, which eliminates the kinetic term in the action \eqref{eq:acao} and simultaneously allows freedom in choosing the functional form of $\varphi(r)$. We then proceed by using Eq.~\eqref{eq:campoescalar} to determine the potential.
\begin{widetext}
\begin{eqnarray}
   && V(r) = \frac{W(r)}{2 \kappa ^2}  \Biggl\{\int \left[\frac{1}{A(r) B(r)^3 \Sigma (r)^2 W(r)^2} \right] \Biggl[ B(r)^2 \biggl(\Sigma '(r) \left(\Sigma (r) W(r) A''(r)+A'(r) \left(\Sigma (r)
   W'(r)+3 W(r) \Sigma '(r)\right)\right) 
   	\nonumber \\
   && \qquad -\Sigma (r) W(r) A'(r) \Sigma ''(r)\biggr)+A(r) \biggl(B(r) \bigl(3
   \Sigma (r) W(r) B'(r) \Sigma ''(r)+\Sigma '(r) \bigl(\Sigma (r) W(r) B''(r)+B'(r) \bigl(3 W(r) \Sigma
   '(r) 
   	\nonumber \\ 
   && \qquad
   -\Sigma (r) W'(r)\bigr)\bigr) 
   +2 B(r) \left(W'(r) \left(\Sigma (r) \Sigma ''(r)+\Sigma
   '(r)^2\right)-W(r) \left(\Sigma (r) \Sigma ^{(3)}(r)+3 \Sigma '(r) \Sigma ''(r)\right)\right)-2 B(r)^2
   W'(r)\bigr) 
   	\nonumber \\ 
   && \qquad \qquad
   -2 \Sigma (r) W(r) B'(r)^2 \Sigma '(r)\biggr) \Biggr] \,
   dr \Biggr\}.
\label{eq:pot}
\end{eqnarray}
\end{widetext}

Given the functional forms of $\mathcal{L}$ in Eq. \eqref{eq:L}, of  $\mathcal{L}_F$ in Eq. \eqref{eq:LF}, and of $V(r)$ in Eq. \eqref{eq:pot}, we can finally use the functions of the Bardeen metric
\begin{equation}
\begin{aligned}
   & A (r) = 1-\frac{2 M r^2}{\left(q_m^2+r^2\right)^{3/2}}, \\
    & B(r) = 1/A(r) ,\\
    & \Sigma(r) = r.
    \end{aligned}
    \label{eq:metricabardeen}
\end{equation}
Now, substituting these values into Eq. \eqref{eq:LF}, we have
\begin{equation}
    \mathcal{L}_F \equiv \frac{d \mathcal{L}}{d F} = \frac{15 M r^6}{\kappa ^2 \left(q_m^2+r^2\right)^{7/2} W(r)},
    \label{eq:principal}
\end{equation}
where we recall that the field strength tensor scalar reads as $F = q_m^2/2r^4$. Our goal is to obtain a coupling function $W(r)$ such that the resulting electromagnetic Lagrangian density is linear, $\mathcal{L}(F) \propto F$. To do this, we equate in the above relation
\begin{equation} \label{eq:condf}
  \frac{15 M r^6}{\kappa ^2 \left(q_m^2+r^2\right)^{7/2} W(r)} =  F(r)^n,
\end{equation}
where $n$ is a natural number (for a more complete description in which $n$ can be negative numbers, see \cite{RBHLE}). 

\begin{figure}[t!]
    \centering
    \includegraphics[scale=0.7]{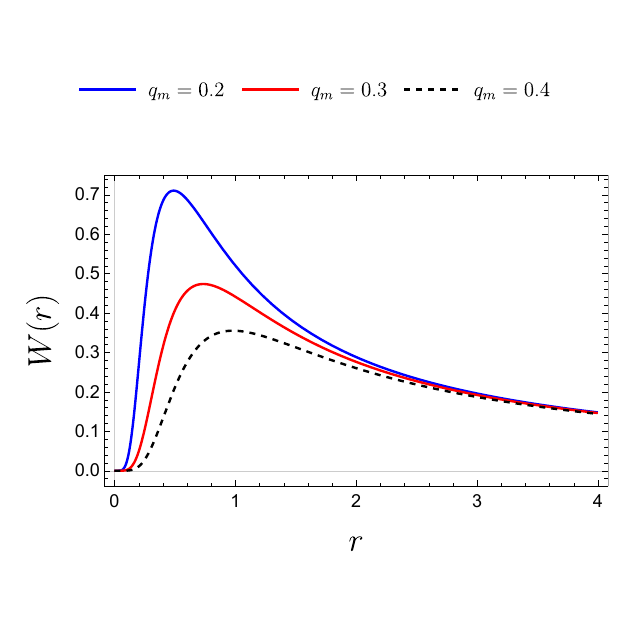}
    \caption{The coupling function $W(r)$ given by Eq.\eqref{eq:wmr}, for $M=1$ and $n=0$.}
    \label{fig:wm(r)1}
\end{figure}

Furthermore, with this choice we find the function $W(r)$ to be
\begin{equation}
    W(r) =  \frac{15 M 2^{n-3} q_m^{-2 n} r^{4 n+6}}{\pi  \left(q_m^2+r^2\right)^{7/2}},
    \label{eq:wmr}
\end{equation}
in terms of the radial coordinate and the parameters of the model. To see what this implies, in Fig. \ref{fig:wm(r)1} we show the behavior of this coupling function in relation to the radial coordinate $r$ for three different values of the magnetic charge and for $n=0$.

The above expresion for the coupling function $W(\varphi)$ allow us to to find the Lagrangian density \eqref{eq:L} and the potential \eqref{eq:pot} as
\begin{equation}
    \mathcal{L}(r) =  \frac{2^{-n-1} q_m^{2 n+2} r^{-4 (n+1)}}{n+1},
    \label{eq:Lrm}
\end{equation}
and
\begin{equation}
    V(r) = \frac{3 M q_m^2 \left(4 (n+1) q_m^2+(4 n-1) r^2\right)}{2 \kappa ^2 (n+1)
   \left(q_m^2+r^2\right)^{7/2}},
   \label{eq:V(r)}
\end{equation}
respectively. It is immediately seen that the Lagrangian density \eqref{eq:Lrm} is different from that one proposed by Beato and Garcia \eqref{eq:LrBardeen}. In order to write it in terms of the field strength scalar $F$ we can use its dependence with the radial coordinate $r$ (valid  only for $q_m >0$), that is,
\begin{equation}
  r=  \sqrt[4]{\frac{q_m^2}{2F }},
\end{equation}

\begin{figure}[t!]
    \centering
    \includegraphics[scale=0.7]{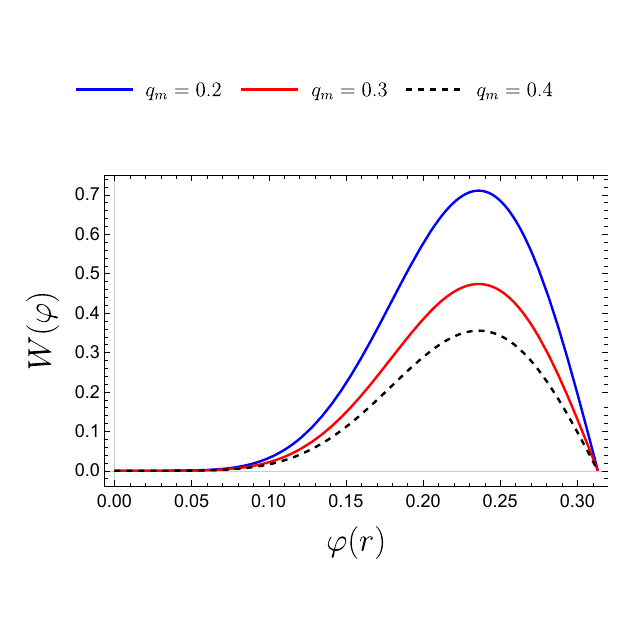}
    \caption{The coupling function $W(\varphi)$ given by Eq.\eqref{eq:wmphi}, for $M=1$ and $n=0$.}
    \label{fig:wm(phi)1}
\end{figure}
\begin{figure}[t!]
    \centering
    \includegraphics[scale=0.7]{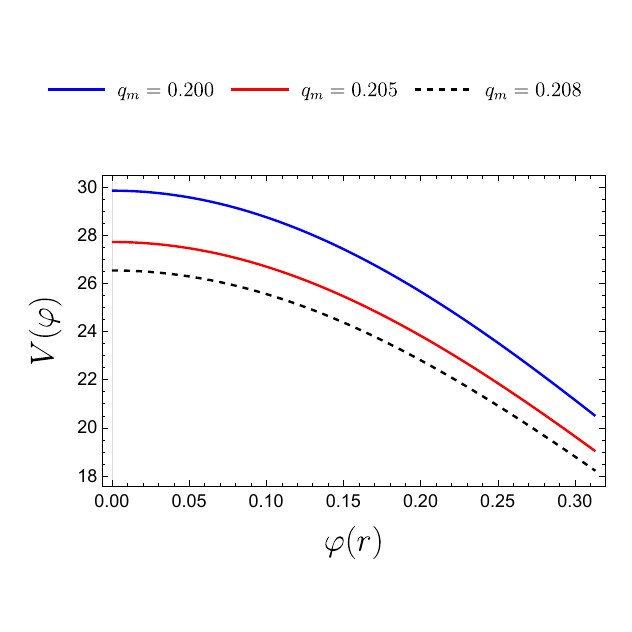}
    \caption{The potential $V(\varphi)$ given by Eq.\eqref{eq:Vmphi}, for $M=1$ and $n=0$.}
    \label{fig:Vm(phi)1}
\end{figure}

Now, we write a function  $\mathcal{L}(F)$
\begin{equation}
    \mathcal{L}(F) = \frac{F^{n+1}}{n+1},
    \label{eq:Lmagnetico}
\end{equation}
and attempt to pick a linear Lagrangian, that is, $n=0$ above, and check the equations of motion for the coupling and scalar potential that supports the consistence of such a pick. To this end, we first recall that, as mentioned above, we have freedom of choice of the scalar field function. Using the choice 
\begin{equation}
    \varphi(r) = \frac{1}{ \kappa }\,\tan ^{-1}\left(\frac{r}{q_m}\right),
    \label{eq:phi(r)}
\end{equation}
which is a functional form commonly used in the literature to describe the scalar field \cite{Rodrigues:2023vtm,Alencar:2024yvh}.
This function tends to zero when $r \rightarrow 0$ and to $\pi/2 \kappa$ when $r \rightarrow \infty$, and is therefore bounded between this interval. We can invert Eq. \eqref{eq:phi(r)}, and then substitute the result into Eqs. \eqref{eq:V(r)} and \eqref{eq:wmr} to write
\begin{equation}
    W( \varphi) = \frac{15 M 2^{n-3} q_m^{2 n-1} \tan ^{4 n+6}\left( \kappa  \varphi \right)}{\pi  \left( \sec \left( \kappa  \varphi
   \right)\right)^{7}},
    \label{eq:wmphi}
\end{equation}
and
\begin{equation}
    V( \varphi) = \frac{3 M \cos ^5\left( \kappa  \varphi \right) \left(5 \cos \left(2 \kappa  \varphi \right)+8 n+3\right)}{4 \kappa ^2 (n+1) q_m^3}.
    \label{eq:Vmphi}
\end{equation}
for the coupling function and the scalar potential, respectively.

We can readily verify that for $n=0$, the function $W(\varphi)$ given in Eq.~\eqref{eq:wmphi} depends inversely on $q_m$, i.e., an increase in the magnetic charge leads to a decrease in the value of the coupling function. This behavior is illustrated in Fig.~\ref{fig:wm(phi)1}. Furthermore, we observe that $W(\varphi)$ tends to zero in both limits of $r$, namely $r \to 0$ and $r \to \infty$.  
We also depict the behavior of the potential $V(\varphi)$ in Fig.~\ref{fig:Vm(phi)1} for the values $q_m = 0.200$, $0.205$, and $0.208$, noting that an increase in the charge generally decreases the value of the potential. Therefore, these functions are smooth for $n=0$, confirming that the Bardeen solution can be obtained using linear electrodynamics as the matter source, with consistent expressions for both the coupling function and the scalar field. We now turn our attention to the electrically charged case.

\subsection{Electric Bardeen solution}\label{sec:doisb}

As discussed in Ref. \cite{regular5}, Bardeen's solution can also be obtained by considering a purely electric tensor in the framework of NEDs. Using our framework, a purely electric solution has, in the line element \eqref{eq:ds}, as its only non-vanishing component the following
\begin{equation} \label{eq:electricc}
    F^{10} = - F^{01} = \frac{q_eA(r)B(r)}{\Sigma^2(r) \mathcal{L}_F(r) W(\varphi)}, 
\end{equation}
which is slightly different from the form presented in \cite{regular5} as a result of the non-minimal coupling given the fact that we have the presence of the function $W(\varphi)$ in the definition of the electric field. In our case, Einstein's equations read as
\begin{widetext}
\begin{eqnarray}
   && \frac{\Sigma (r) B'(r) \Sigma '(r)-B(r) \left(2 \Sigma (r) \Sigma ''(r)+\Sigma '(r)^2\right)+B(r)^2}{B(r)^2
   \Sigma (r)^2}
   \nonumber \\ 
   && \qquad \qquad
   -\kappa ^2 \left(\frac{q_e^2 A(r) B(r)}{\mathcal{L}_F(r) \Sigma (r)^4 W(r)}+\frac{
   \epsilon (r) \varphi '(r)^2}{B(r)} 
   +\mathcal{L}(r) W(r)+ V(r)\right) = 0,
   \label{eq:00E}
\end{eqnarray}
\begin{eqnarray}
        \frac{\Sigma (r) A'(r) \Sigma '(r)+A(r) \left(\Sigma '(r)^2-B(r)\right)}{A(r) B(r) \Sigma
   (r)^2}+\kappa ^2 \left(\frac{q_e^2 A(r) B(r)}{\mathcal{L}_F(r) \Sigma (r)^4 W(r)}-\frac{ \epsilon
   (r) \varphi '(r)^2}{B(r)}+\mathcal{L}(r) W(r)+ V(r)\right) =0,
 \label{eq:11E}
\end{eqnarray}
\begin{eqnarray}
    &&    \frac{A(r) B'(r) \left(\Sigma (r) A'(r)+2 A(r) \Sigma '(r)\right)+B(r) \left(\Sigma (r) \left(A'(r)^2-2 A(r)
   A''(r)\right)-2 A(r) \left(A'(r) \Sigma '(r)+2 A(r) \Sigma ''(r)\right)\right)}{4 A(r)^2 B(r)^2 \Sigma
   (r)}
   	\nonumber \\ 
   && \qquad \qquad \qquad
   -\kappa ^2 \left(\frac{ \epsilon (r) \varphi '(r)^2}{B(r)}+\mathcal{L}(r) W(r)+ V(r)\right) =0,
    \label{eq:22E}
\end{eqnarray}
\end{widetext}
respectively. Note that while these equations are only slightly different from the purely magnetic case,  this is however sufficient to also modify the form of the Lagrangian density and its derivative and the potential. To work their explicit forms out, we combine Eqs. \eqref{eq:00E} and \eqref{eq:22E} to write
\begin{widetext}
\begin{eqnarray}
    \mathcal{L} &=& \frac{1}{4 \kappa ^2
   A(r)^2 B(r)^2 \Sigma (r) W(r)}
\Biggl\{B(r) \Sigma (r) A'(r)^2+A(r) \left(A'(r) \left(\Sigma (r) B'(r)-2 B(r) \Sigma '(r)\right)-2 B(r) \Sigma (r) A''(r)\right)
		\nonumber \\ 
	&& \qquad \qquad -2 A(r)^2 \left(2
   B(r) \left( \kappa ^2 \Sigma (r) \left(B(r) V(r)+\varepsilon (r) \varphi '(r)^2\right)+\Sigma ''(r)\right)-B'(r) \Sigma '(r)\right)\Biggr\},
     \label{eq:leletrico}
\end{eqnarray}
\begin{eqnarray}
        \mathcal{L}_F &=& 4 \kappa ^2 q_e^2 A(r)^3 B(r)^3  \Biggl\{\Sigma (r)^2 W(r) \biggr[-B(r) \Sigma (r)^2 A'(r)^2+A(r) \Sigma (r) \bigl(2 B(r) \Sigma (r)
   A''(r)+A'(r) \bigl(2 B(r) \Sigma '(r) 
   		\nonumber \\ 
   && \qquad  -\Sigma (r) B'(r)\bigr)\bigr)
   +2 A(r)^2 \left(\Sigma (r) B'(r) \Sigma '(r)-2 B(r) \left(\Sigma (r)
   \Sigma ''(r)+\Sigma '(r)^2\right)+2 B(r)^2\right)\biggl]\Biggr\}^{-1},
    \label{eq:lfeletrico}
\end{eqnarray}
\begin{eqnarray}
    V(r) &=& \frac{W(r)}{4 \kappa ^2}  \Biggl\{\int \left[\frac{1}{A(r)^2 B(r)^3 \Sigma (r)^2 W(r)^2\Sigma(r)^2} \right] \Biggl[A(r) B(r) \bigl(\Sigma (r)^2 \left(-A'(r)\right) B'(r) W'(r)+6 A(r) W(r) B'(r) \Sigma '(r)^2
  	\nonumber	\\ 
  && \qquad +2 A(r) \Sigma
   (r) W(r) \left(B''(r) \Sigma '(r)+3 B'(r) \Sigma ''(r)\right)\bigr)+B(r)^2 \biggl(6 A(r) W(r) \Sigma '(r)
   \left(A'(r) \Sigma '(r)-2 A(r) \Sigma ''(r)\right)  
   		\nonumber 	\\
 && \qquad  +2 A(r) \Sigma (r) \left(W(r) \left(A''(r) \Sigma
   '(r)-2 A(r) \Sigma ^{(3)}(r)\right)+A'(r) \left(2 \Sigma '(r) W'(r)-W(r) \Sigma
   ''(r)\right)\right)
   		\nonumber \\
  && \qquad +\Sigma (r)^2 \left(-W'(r)\right) \bigl(A'(r)^2 
  -2 A(r) A''(r)\bigr)\biggr)-4 A(r)^2
   \Sigma (r) W(r) B'(r)^2 \Sigma '(r) \Biggr] \,
   dr \Biggr\}.
\label{eq:pote}
\end{eqnarray}
\end{widetext}
Now we can use the resemblances with the previous case so that by substituting these results into Eq.~\eqref{eq:11E} we get again Eq.~\eqref{eq:epsilon}. The metric function is also the same, but with the replacement $q_m \rightarrow q_e$. 

\begin{figure}[t!]
    \centering
    \includegraphics[scale=0.7]{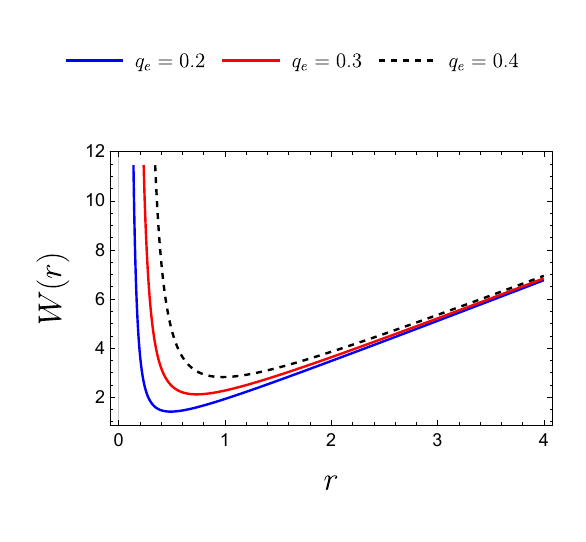}
    \caption{The coupling function $W(r)$ given by Eq.\eqref{eq:wre} for the values: $n=0, M=1$.}
    \label{fig:wer}
\end{figure}

To find the explicit matter sources, we shall use the same strategy as before, that is, we impose that $\mathcal{L}_F =  F(r)^n$ to find a function $W(r)$ that generates a linear Lagrangian density via the condition
\begin{equation}
    \mathcal{L}_F = \frac{d \mathcal{L}}{d F} = \frac{\kappa ^2 \left(q_e^2+r^2\right)^{7/2}}{15 M r^6 W(r)} =  F(r)^n .
\end{equation}
Since the electric field scalar reads now
\begin{equation}
    F (r) = -\frac{225 M^2 q_e^2 r^8}{2 \kappa ^4 \left(q_e^2+r^2\right)^7},
    \label{eq:Feletrico}
\end{equation}
we find
\begin{equation}
  W(r) =  \frac{(-2)^n\kappa ^{4 n+2}\left(q_e^2+r^2\right)^{7
   n+\frac{7}{2}}}{15^{2 n+1} M^{2 n+1}  q_e^{2 n} r^{8 n+6}}.
   \label{eq:wre}
\end{equation}
This function is depicted in Fig.~\ref{fig:wer}. 
With the coupling function in place, we can solve for 
\begin{equation}
   \mathcal{ L}(r) = -\frac{ M^{2 n+2}  q_e^{2 n+2} r^{8
   n+8} }{\left(q_e^2+r^2\right)^{+7 n+7}\kappa ^{4 n+4}(n+1)\left(\frac{2}{225}\right)^{n+1} e^{i \pi  n}}.
   \label{eq:Lele}
\end{equation}
Then, if we take $n=0$ in the above equation we get
\begin{equation}
     \mathcal{ L}(r)|_{n=0} = -\frac{225 M^2 q_e^2 r^8}{2 \kappa ^4 \left(q_e^2+r^2\right)^7},
\end{equation}
which is exactly the same expression as the one for the scalar $F(r)$ in Eq.~(\ref{eq:Feletrico}). So we conclude that the Lagrangian density can indeed be consistently made a linear function of the Maxwell scalar when choosing $n=0$. It is worth emphasizing that due to the linearity of the Lagrangian density  in this case, the problem of the multivaluedness which typically affects the framework or regular black holes in electrically charged NED models is avoided, which is therefore an obvious advantage of our approach.

\begin{figure}[t!]
    \centering
    \includegraphics[scale=0.7]{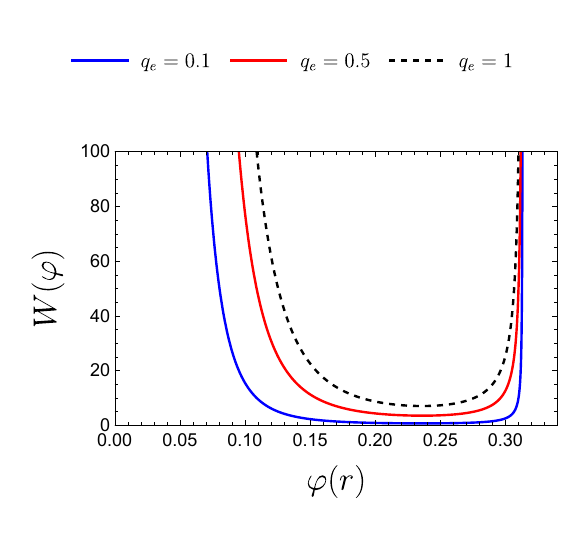}
    \caption{The coupling function $W(\varphi)$ given by Eq.\eqref{eq:wphie} for the values: $n=0, M=1$.}
    \label{fig:wphie}
\end{figure}

\begin{figure}[t!]
    \centering
    \includegraphics[scale=0.7]{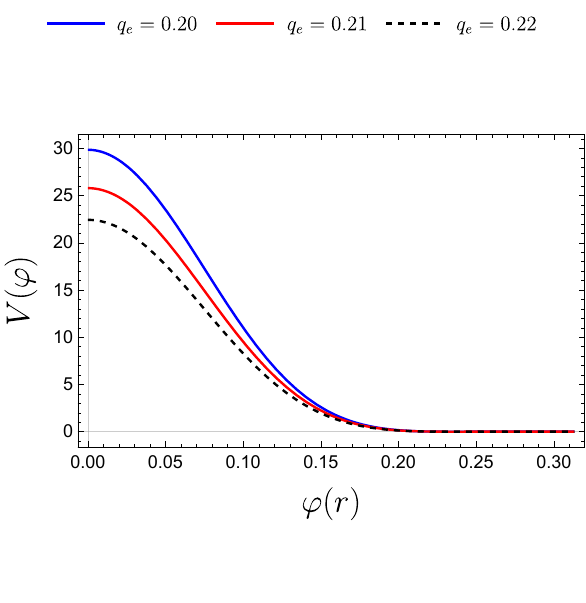}
    \caption{The scalar field potential $V(\varphi)$ given by Eq.\eqref{eq:Vphie}, for the values: $n=0, M=1$.}
    \label{fig:vphie}
\end{figure}

Moving on, we have from Eq.~\eqref{eq:pot} the expression
\begin{equation}
  V(r) =   \frac{12 M (n+1) q_e^4-3 M (6 n+1) q_e^2 r^2}{2 \kappa ^2 (n+1) \left(q_e^2+r^2\right)^{7/2}},
  \label{eq:v(r)e}
\end{equation}
such that by using Eq.~\eqref{eq:phi(r)} again, we can write the coupling function and the potential as a function of the scalar field as
\begin{equation}
    W ( \varphi) =   \frac{(-2)^n  \kappa ^{4 n+2}\left(q_e^2 \sec ^2\left( \kappa  \varphi \right)\right)^{7 n+\frac{7}{2}}}{15^{2 n+1} M^{2 n+1} q_e^{2 n} \left(q_e \tan \left( \kappa  \varphi
   \right)\right)^{8 n+6}}
   ,
   \label{eq:wphie}
\end{equation}
and
\begin{equation}
    V( \varphi) = -\frac{3 M \cos ^4\left( \kappa  \varphi \right) \left((6 n+1) \tan ^2\left( \kappa  \varphi
   \right)-4 (n+1)\right)}{2 \kappa ^2 (n+1) \left(q_e^2 \sec ^2\left( \kappa  \varphi
   \right)\right)^{3/2}},
   \label{eq:Vphie}
\end{equation}
respectively. Note that, as stated before, the scalar field is limited between the values  $0 \leq \varphi \leq \pi/2 \kappa$, which represent the limit values of the radial coordinate. Using this information in Eq.~\eqref{eq:wphie} we find
\begin{equation}
   \lim_{\varphi \rightarrow 0} W(\varphi) = \lim_{\varphi \rightarrow \pi/2 \kappa} W(\varphi) \to \infty.
\end{equation}
These divergences obviously raise questions about the validity of the model. However, it is important to note that this coupling function, although associated with several experimentally measurable physical quantities, is not a direct observable itself. Therefore, to verify the validity of the model, we shall calculate four physical quantities of interest: the energy density $\rho$, the radial pressure $p_r$, the tangential pressure $p_t$, besides the electric field $E$ itself. These objects read, for a general metric, as
\begin{eqnarray}
    \rho (r) = T^{0}_{\ 0} &=& \frac{q_e^2 A(r) B(r)}{\mathcal{L}_F(r) \Sigma (r)^4 W(r)}+\frac{\varepsilon (r) \varphi'(r)^2}{B(r)} 
   		\nonumber \\ 
   	&&+\mathcal{L}(r) W(r)+ V(r),
\end{eqnarray}
and
\begin{eqnarray}
   p_r (r) = -T^{1}_{\ 1} &=& \frac{ \varepsilon (r) \varphi
   '(r)^2}{B(r)}- \frac{q_e^2 A(r) B(r)}{\mathcal{L}_F(r) \Sigma (r)^4 W(r)} 
   		\nonumber \\ 
   && - \mathcal{L}(r) W(r)- V(r),
\end{eqnarray}
and
\begin{equation}
    p_t(r) = -T^{2}_{\ 2} =  - \left[ \frac{ \varepsilon (r) \varphi '(r)^2}{B(r)} + \mathcal{L}(r) W(r) +  V(r) \right],
\end{equation}
respectively, while $E(r)$ is given by Eq.~(\ref{eq:electricc}). 

For the Bardeen solution, we have $\varepsilon (r)= 0$, which leads to $\rho = - p_r$.
Substituting  the metric \eqref{eq:metricabardeen}, the potential \eqref{eq:v(r)e} and the Lagrangian density \eqref{eq:Lele} into the above equations, we obtain 
\begin{equation}
     \rho (r) = - p_r(r)=  \frac{6 M q_e^2}{\kappa^2  \left(q_e^2+r^2\right)^{5/2}},
     \label{eq:rhoe}
\end{equation}
\begin{equation}
    p_t(r) = \frac{ M q_e^2 \left(9 r^2-6 q_e^2\right)}{\kappa^2  \left(q_e^2+r^2\right)^{7/2}},
    \label{eq:pte}
\end{equation}
\begin{equation}
    E(r) = \frac{15 M q_e r^4}{\kappa^2 \left(q_e^2+r^2\right)^{7/2}},
\end{equation}
respectively.

With these expressions, we can verify that all these physical quantities are everywhere finite. We therefore conclude that the divergences in the function $W(\varphi)$ are not propagated to any physical quantity of interest. Fig.~\ref{fig:wphie} shows, for three values of electric charge, the smoothness of this function outside the boundary of its domain. Furthermore, we highlight, in Fig. \ref{fig:vphie}, the behaviour of $V(\varphi)$ for three values of the electric charge. 

It is worth mentioning that Eqs.~\eqref{eq:rhoe} and \eqref{eq:pte} coincide with those reported in \cite{regular5}. One implication of this is that, since for the NED description considered in \cite{regular5} satisfies the condition $p_t = \mathcal{L}_{NED}$, we can make the following comparison
\begin{equation}
    \mathcal{L}_{NED}  = \mathcal{L}W +V.
\end{equation}
This can be understood as the recovery of the usual (non-linear) Lagrangian function, $\mathcal{L}_{NED}$, by adding the potential $V(r)$ together with the product $\mathcal{L}(r)W(r)$ out of the coupling function. This is achieved without even considering $n=0$ in the expressions, and is also valid for the purely magnetic case.

\section{Conclusion \label{sec:conclu}}

In this paper we have re-interpreted the Bardeen regular black hole solution as sourced by a linear electromagnetic field which interacts with a scalar field $\varphi$. This formalism is valid for both purely magnetic and purely electric descriptions. 
The resulting Lagrangian density differs from that presented Ayón-Beato and García \cite{Beato} and depends on an additional arbitrary parameter $n$ encoding a power-law NED Lagrangian, $\mathcal{L}(F) \approx F^{n+1}$. The solution reduces to the linear Maxwell case only when $n=0$; otherwise, it remains within the framework of NED. We furthermore argued that this approach can be applied to other regular solutions, as illustrated in Appendix \ref{sec:ap1} for the Ayón-Beato and García solution, and that it offers advantages over the standard description based solely on NED sources, such as the recovery of the Maxwell weak-field limit and absence of multivaluedness issues on its definition.

To implement our procedure, we began by revisiting the original derivation of Bardeen’s solution with purely magnetic and electric sources, emphasizing the nonlinear character of the associated Lagrangian densities. We also analyzed the Kretschmann scalar, showing its finiteness across space-time and confirming the regularity of the solution. We then introduced Einstein’s equations coupled to electrodynamics and a scalar field with non-minimal coupling between each another as given by a function $W(\varphi)$ such that the usual minimally coupled case is recovered by setting $W(\varphi)=1$. In the sequel we assumed a very general static and spherically symmetric metric, to construct our framework separately for purely magnetic and purely electric solutions.

For the magnetic case, we derived the equations of motion, the Lagrangian density, and its derivative, together with an additional relation connecting the scalar field and the parameter  $\varepsilon(\varphi)$ determining the canonical or phantom nature of its kinetic field. We found that $\varepsilon=0$ under the conditions $B(r)=1/A(r)$ and $\Sigma^{\prime\prime}=0$, which are satisfied by the Bardeen solution. In this situation, the scalar kinetic term vanishes. We then determined the potential $V(r)$ and the functional form of $W(r)$ from the field equations. Finally, assuming an expression for the electromagnetic Lagrangian density of the form $\mathcal{L}_F = F^n$ we proof that $n=0$  (i.e. a linear electrodynamics Lagrangian) can be chosen while keeping the field equations consistent. Our analysis shows that the coupling function and the potential are real, positive, and regular for all $r$, further supporting the consistency of the construction. This provides a framework in which Bardeen’s solution can be interpreted either through NED alone or via a combination of Maxwell electrodynamics plus a non-minimally coupled scalar field.

We extended the construction to the purely electric case, where the electromagnetic tensor structure differs from the magnetic case. Although the equations of motion and the resulting Lagrangian densities acquire new forms, the key features of the framework remain: the kinetic term vanishes, $\varepsilon=0$, the scalar field can still be freely chosen, and Maxwell electrodynamics is recovered when $n=0$. The potential $V(\varphi)$ retains the same regular and positive behavior, while the coupling function $W(\varphi)$ diverges at the boundaries of the range. Importantly, this divergence does not translate into divergences in physically relevant quantities such as energy density, pressure, or the electric field. We explicitly showed that the action and Einstein’s equations remain well behaved, and that the matter content reproduces the same energy–momentum components as in \cite{regular5}. For Bardeen’s solution, the relation $\mathcal{L}_{NED} = \mathcal{L}W + V$ follows naturally. Furthermore, the linearity of the Lagrangian ensures the absence of the multivalued structure typically present in electrically charged NED theories.

The results found in this paper hinge around the problem of finding field sources supporting ad-hoc constructed solutions that implement desired features in contrast to the canonical method of setting an action, apply variational procedures to find the corresponding field equations, and solve them under certain conditions for the symmetries of the problem and the nature of the matter fields. Our proposal puts forward that there is not a single way of interpreting any given such solutions, and that different frameworks will come with their advantages and disadvantages over the others that need to be uncovered to get some insights upon the features of the solutions and the corresponding theories. 

In the present case, our formalism bypasses the problem of not having the usual Maxwell weak-field limit in the electrodynamics sector, given the pool of experiments supporting that this is indeed the case. Other problems that can be mentioned associated to the usual NED description of these solutions include instabilities in the effective metric description, modifications of the canonical black hole shadows description, additional constraints required for causality, unitarity and stability, and the lack of a consistent thermodynamic framework. On the other hand, non-minimal couplings between different matter sectors of the theory may bring their own problems regarding e.g. an enhanced complexity of the corresponding field equations, or the potential existence of fifth forces and non-geodesic behaviour.

It is worth pointing out that it is not clear whether it is always possible to find matter sources for every regular solution within our non-minimal coupling approach yet. Therefore, beyond specific cases (like the two considered here for Bardeen in the main text, and for Ayón-Beato and García in the Appendix), it would be interesting to establish a general framework for every regular black hole geometry derived from the action considered here. This would facilitate the comparison of approaches, their corresponding advantages and drawbacks, and to set light on the degeneracy of action building blocks supporting regular space-time geometries. Work along this path is currently underway.

\section*{Acknowledgements}

MER thanks Conselho Nacional de Desenvolvimento Cient\'ifico e Tecnol\'ogico - CNPq, Brazil, for partial financial support. This study was financed in part by the Coordena\c{c}\~{a}o de Aperfei\c{c}oamento de Pessoal de N\'{i}vel Superior - Brasil (CAPES) - Finance Code 001.
FSNL acknowledges support from the Funda\c{c}\~{a}o para a Ci\^{e}ncia e a Tecnologia (FCT) Scientific Employment Stimulus contract with reference CEECINST/00032/2018, and funding through the research grants UIDB/04434/2020, UIDP/04434/2020 and PTDC/FIS-AST/0054/2021.
DRG is supported by the Agencia Estatal de Investigación Grant Nos. PID2022-138607NB-I00 and CNS2024-154444, funded by MICIU/AEI/10.13039/501100011033 (Spain).

\appendix
\section{Ayón-Beato and García (ABG) solution \label{sec:ap1}}

This metric was presented in \cite{Ayon-Beato:1999kuh} as an exact solution for an electrically charged regular black hole with a NED source. It corresponds to Maxwell theory in the limiting case of weak fields and has two horizons for $|q| < 1.05m$. Here, we will describe this solution as a purely magnetic solution with $q=q_m$ thus being the magnetic charge. The reason  behind this choice is that the electric NED Lagrangian density cannot always be represented in a analytic form $\mathcal{L}(F)$. For example, in \cite{Ayon-Beato:1999kuh} they use the Legendre transformation $\mathcal{H} = 2 F \mathcal{L}_F - \mathcal{L}$ to work in the so-called $P$ framework instead of the usual $F$. Without a complete analytic form in the $F$ framework, the Lagrangian density is a multivalued function of $F$, since one value of $F$ can correspond to more than one possible value of $\mathcal{L}_F$ (see \cite{Bronnikov:2000vy} for more explanation of the preference for magnetic solutions).

The ABG metric is given by the functions
\begin{equation}
\begin{aligned}
   & A (r) = 1-\frac{2 M \left(1-\tanh \left(  q_m^2/2M r\right)\right)}{r}, \\
    & B(r) = 1/A(r) ,\\
    & \Sigma(r) = r.
    \end{aligned}
    \label{eq:metricaABG}
\end{equation}
Using this metric in \eqref{eq:LF} we find $\mathcal{L}_F$, and then we impose 
\begin{equation}
    \mathcal{L}_F = \frac{\left(4 M r-q_m^2 \tanh
   \left(\frac{q_m^2}{2 M r}\right)\right)
   \text{sech}^2\left(\frac{q_m^2}{2 M
   r}\right)}{2 \kappa ^2 M r W(r)} = F^{n}.
\end{equation}
We solve this equation to find the coupling function $W$ that leads to a linear Lagrangian density for $n=0$, which turns out to be
\begin{equation}
    W(r) = \frac{2^n \left(\frac{q_m^2}{r^4}\right)^{-n}
   \left(4 M r-q_m^2 \tanh
   \left(\frac{q_m^2}{2 M
   r}\right)\right)}{\kappa ^2 M r \left(\cosh
   \left(\frac{q_m^2}{M r}\right)+1\right)},
\end{equation}
such that this function leads directly to \eqref{eq:Lrm}. It is interesting to note that both Bardeen and ABG solution have the same magnetic Lagrangian density, the differences between them arising at the level of the coupling function and the scalar potential. The latter can be found as (we are again using Eq.\eqref{eq:phi(r)})
\begin{equation}
    V(r)  = \frac{q_m^2 \text{sech}^2\left(\frac{q_m^2}{2 M r}\right) \left(4 M n r+q_m^2 \tanh
   \left(\frac{q_m^2}{2 M r}\right)\right)}{4 \kappa ^2 M (n+1) r^5}.
\end{equation}

\begin{figure}[t!]
	\centering
	\includegraphics[scale=0.7]{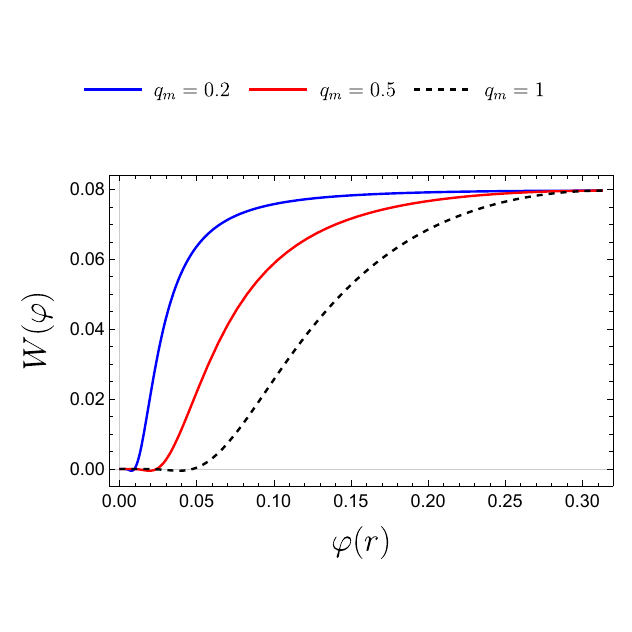}
	\caption{The coupling function $W(\varphi)$ given by Eq.\eqref{eq:wabg} for the values: $n=0, M=1$.}
	\label{fig:wmabg}
\end{figure}
\begin{figure}[t!]
	\centering
	\includegraphics[scale=0.7]{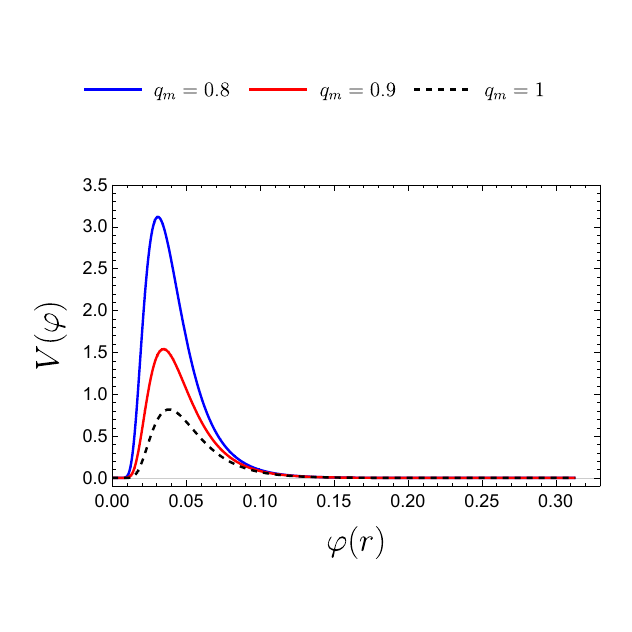}
	\caption{The scalar field potential $V(\varphi)$ Eq.\eqref{eq:vabg} for the values: $n=0, M=1$.}
	\label{fig:vabg}
\end{figure}

Then, we see that 
\begin{equation}
    \mathcal{L}W + V = \frac{q_m^2 }{\kappa^2  r^4}\text{sech}^2\left(\frac{q_m^2}{2 M r}\right).
\end{equation}
We note again that this quantity is independent of $n$ and represents exactly what would be only $\mathcal{L}$ in a description based purely on NED. Finally, we use the connection between the scalar field $\varphi$ and the radial coordinate to write
\begin{eqnarray}
   && W(\varphi) = \frac{2^n \left(\frac{\cot ^4(\kappa  \varphi )}{q_m^2}\right)^{-n} }{\kappa ^2 M \left(\cosh
   \left(\frac{q_m\text{cot}(\kappa  \varphi )}{M}\right)+1\right)} \times 
   		\nonumber \\
  && \qquad  \left[4 M-q_m \cot (\kappa  \varphi ) \tanh
   \left(\frac{q_m \cot (\kappa  \varphi )}{2 M}\right)\right],
   \label{eq:wabg}
\end{eqnarray}
and
\begin{eqnarray}
   && V ( \varphi) = \frac{\cot ^4(\kappa\varphi) \text{sech}^2\left(\frac{q_m \cot (\kappa  \varphi)}{2 M}\right) }{4 \kappa ^2 M (n+1) q_m^2} \times 
   		\nonumber \\
   && \qquad \left[4 M n+q_m \cot (\kappa  \varphi) \tanh \left(\frac{q_m
   \cot (\kappa  \varphi)}{2 M}\right)\right],
   \label{eq:vabg}
\end{eqnarray}
and we represent these functions in the Figs.~\ref{fig:wmabg} and \ref{fig:vabg}. It is clear that both functions are well-defined in the entire domain.




\end{document}